# PhaseMAC: A 14 TOPS/W 8bit GRO based Phase Domain MAC Circuit for In-Sensor-Computed Deep Learning Accelerators

Kentaro Yoshioka, Yosuke Toyama, Koichiro Ban, Daisuke Yashima, Shigeru Maya, Akihide Sai, Kohei Onizuka

Toshiba, Kawasaki, Japan, Email:kentaro3.yoshioka@toshiba.co.jp

## Abstract

PhaseMAC (PMAC), a phase domain Gated-Ring-Oscillator (GRO) based 8bit MAC circuit, is proposed to minimize both area and power consumption of deep learning accelerators. PMAC composes of only digital cells and consumes significantly smaller power than standard digital designs, owing to its efficient analog accumulation nature. It occupies 26.6 times smaller area than conventional analog designs, which is competitive to digital MAC circuits. PMAC achieves a peak efficiency of 14 TOPS/W, which is best reported and 48% higher than conventional arts. Results in anomaly detection tasks are demonstrated, which is the hottest application in the industrial IoT scene.

## Introduction

DNN is a key enabler of new paradigm in cognitive computing, spanning from speech recognition to anomaly detection for industrial systems. Edge-computing with ultra low-power DLAs are essential to adopt DNN into data-traffic, power-supply limited factories; microwatt operation is one of the ultimate goals. It is well known that in CNNs, memory power($E_{Memory}$) is mostly consumed by the partial output movements. On the other hand, for fully-connected (FC) and LSTM models (DNNs served in industrial applications), most $E_{Memory}$ is consumed by readout of weight parameters, which can be 10x larger than the computation energy of digital MACs($E_{DMAC}$). However, by simply batching inputs, weights can be broadcast and reused; significantly cutting down $E_{Memory}$(Fig.1). In other words, batching reduces the effective memory-readouts/inference. For batch size=64, calculated based on ref.[1] and anomaly detection FC models, $E_{DMAC}$ becomes the dominant power source; the computation power is 3 times larger than memory. Therefore, to ultimately scale down the power of DLAs to be operated in batteries or energy-harvesting, power-efficiency of the MAC circuits must be improved. The overhead of 64x batching is increased memory area of only 5% and longer latency, which is acceptable for most applications.

Analog signal processing have been explored to achieve better efficiency than digital MAC(DMAC)[2,3]. However, only 1-3bit available resolution limits the application to image classification, since 6-8bits are required to execute more complex tasks such as speech and sensor time series data analysis today[4]. Moreover, analog MACs has a large area overhead compared to DMACs. In time domain processing[2], since time information cannot be sampled, accumulation is not realized and huge silicon area is required. Switched-capacitor (SC) based MAC[3] also achieve efficient computing but consumes plentitude of area-consuming analog elements and its area is >20x larger than DMACs.

## PhaseMAC

We propose the PhaseMAC(PMAC), which is >20x area-efficient than conventional analog MACs and 8x power-efficient than DMACs. For DLAs whom power is dominated by $E_{DMAC}$(as in Fig.1), PMAC will reduce 66% power of the entire DLA. Fig.2 shows the key concept of the proposed phase domain operation of PMAC. GRO is basically composed of a ring oscillator with a power gating switch; the oscillation is conducted only when the switch is shorted, and when the switch opens, the phase information is saved[5]. A DTC generates a pulse (DTCOUT) proportional to the input signal ($D_{in}$), which is applied to every GRO gating switch, and weight signal ($W$) controls the GRO frequency to enable the "multiply" operation.

Here, 2 sequences of the PMAC operation and its readout is explained. Since $D_{in}$=3, $W$=1 is given at Seq.1, DTCOUT pulse width is $3*t_{inv}$ ($t_{inv}$ is the GRO inverter delay at $W$=1) and the GRO advances its phase for 3 inverter cells, or $0.6\pi$. After DTCOUT sets down, the GRO gating switch simultaneously opens and saves the phase state. Notice that since the oscillation of Seq.2 starts from the advanced phase state, phase domain accumulation is realized. Since the GRO is a 5-stage ring oscillator, the phase state returns to the initial condition after the phase excesses $2\pi$; this event is caught with a counter circuit which counts up when the GRO's phase returns to 0. Therefore, the PMAC can continuously operate as a phase domain accumulator as long as the counter does not saturate. Lastly, the readout logic samples the counter output and the GRO phase as the MSB and LSB, respectively. The GRO phase is quantized by latching each inverter output.

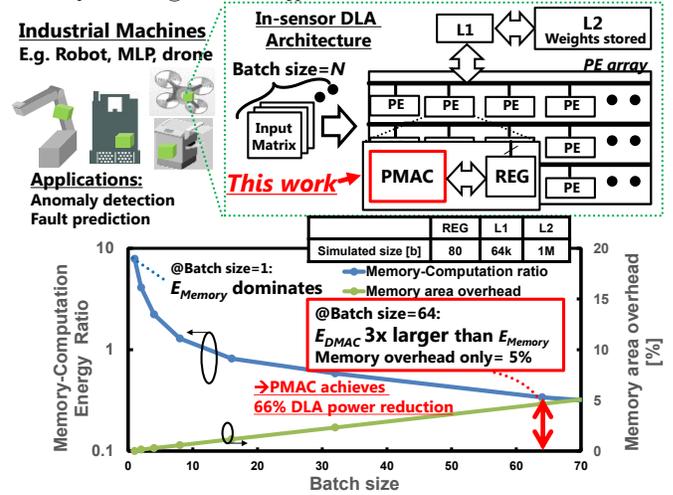

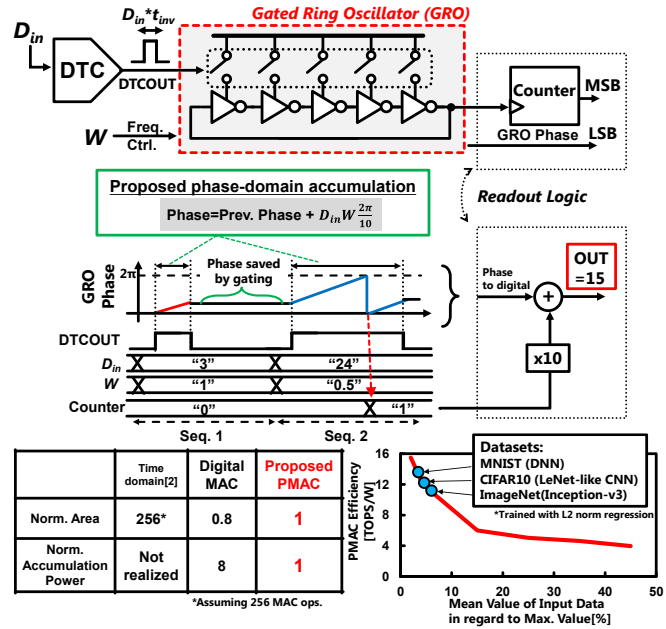

**Fig.1 Batch size versus memory-computation energy ratio**

**Fig.2 Concept of the PhaseMAC**

The PMAC power is mostly cheap inverter transitions where the number corresponds to the $D_{in}$ and $W$ values, (in contrast with the over 40 DFFs in an 8bit DMAC) hence, the power consumption is fundamentally low. Importantly, the PMAC's power consumption

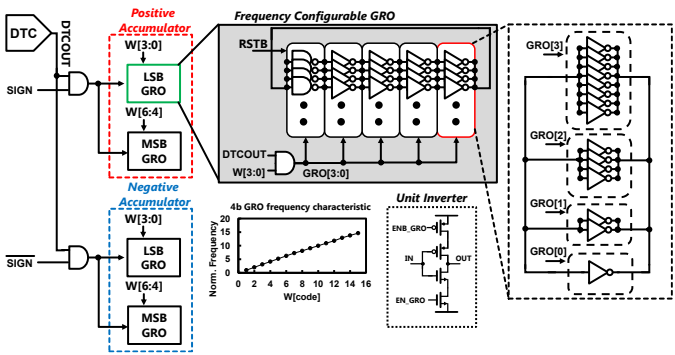

**Fig.3 Block diagram of the Multi-GRO architecture supporting 8b multiplication function.**

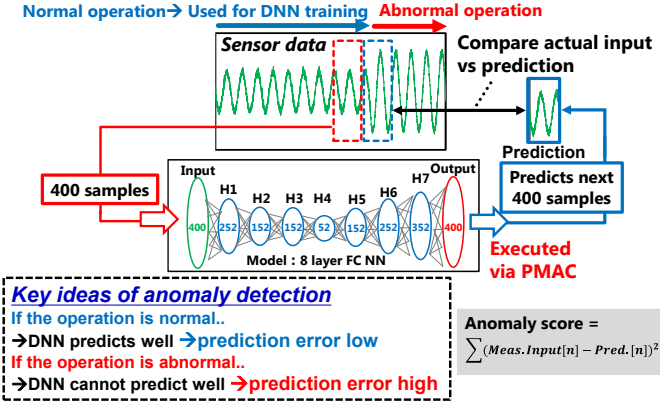

**Fig.4 In-Sensor-Edge-Computing for Anomaly Detection of Industrial Machines**

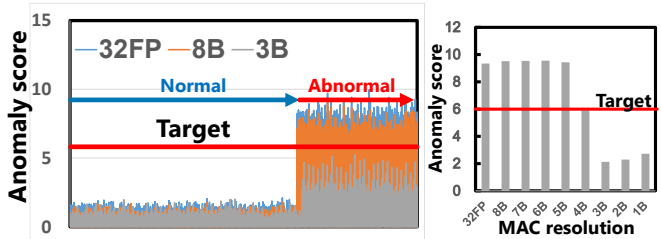

**Fig.5 Measured anomaly score with various MAC resolution**

**TABLE. 1 MNIST classification results**

|  | 32FP | 8b PMAC |
|---|---|---|
| Validation Results[%] | 98.2 | 98.1 |

Model: 5layer FC

**TABLE 2. Performance comparison**

|  | This work | | [2] | [3] |
|---|---|---|---|---|
| Domain | Phase | | Time | Charge |
| Process | 28nm | | 65nm | 40nm |
| Resolution | 8 | | 1 | 3 |
| MAC Area [um²] | 1200 | | 13000* | 12000 |
| MAC Area/Bit [um²] | 150 | | 13000* | 4000 |
| Application | MNIST | Anomaly Detection | MNIST | CIFAR10 |
| Power [uW] | 152 | 170 | N.A. | 228** |
| MAC rate | 780 MHz | 700 MHz | N.A. | 1 GHz |
| Efficiency [TOPS/W] | 14 | 11.6 | 77 | 8.77 |
| Efficiency [TOPS/W*Bit] | 112 | 92.8 | 77 | 26.3 |

*Assuming 256 MAC ops. **Includes memory power

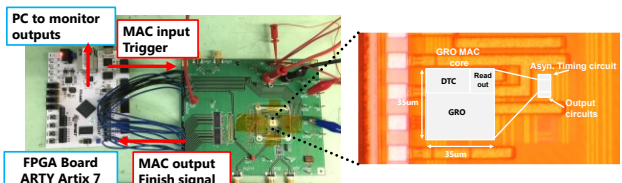

**Fig.6 Chip photo and measurement setup**

correlates with $D_{in}$, $W$ mean values and it is ubiquitous that in DNNs the $D_{in}$, $W$ mean values are very low (3~10%). Therefore, PMAC becomes up to 8x more efficient than DMAC for such operations; PMAC is an optimal MAC circuit for DNNs (Fig.2).

Fig.3 shows the PMAC schematic. To achieve an 8b multiplication function with minimum area, we propose a multi-GRO architecture. Foremost, it is essential to deal with negative numbers for MAC operation in DNNs. To realize such MAC circuits with minimum overhead, we simply utilize a set of two GROs: one GRO to accumulate positive numbers and other to accumulate the negative, the final MAC output is obtained by conducting subtraction in the digital domain. The positive accumulator is activated when the SIGN signal (XNOR of $D_{in}$ and $W$ MSBs) is 1, and vice versa. Moreover, design of GRO with 7b linear frequency control capability is challenging to achieve without calibration. To evade such complexity, 7b frequency resolution is functionally realized by assigning two individual GROs to 3b MSB and 4b LSB, respectively. The readout circuit generates the MAC output by summing the bit shifted MSB GRO output to the LSB GRO output. Inverters turned-off simply act as load capacitors, and hence, the GRO frequency is linearly controlled with $W$ and achieves phase domain multiplication with 7b resolution.

*Measurement results*

PMAC proof of concept chip was fabricated in 28nm CMOS. Anomaly detection task is demonstrated in Fig.4, where the Matrix Multiply is executed by the PMAC, and the host FPGA controls the entire DNN data flow(Fig.6). Unlike ref.[2,3], all 8 DNN layers are computed via analog MAC. In-sensor, ultra low-power anomaly detection is essential for industrial IoT applications, such as motor vibration monitoring in huge plants under strong radio communication and power supply constraints. The autoencoder based DNN inputs 400 points of sensor data and trained to predict the next 400 data points[6]. The system determines that the target is operating normally when the prediction error is small, since the operation is close to the trained normal condition. On the other hand, alerts are made when the prediction error exceeds the threshold; the operation is abnormal and clearly differing from the trained condition. Fig.5 shows the measured anomaly detection task results, where detection with 8bit resolution obtains the anomaly score close to 32FP processing, but less than 4bit quantization clearly degrades the detection accuracy; conventional low resolution analog MAC is not acceptable for this task. Also, MNIST has been tested with a similar FC NN, achieving a TOP-1 validation accuracy of 98.1%, where the 32FP results gave 98.2%(TABLE.1).

TABLE.2 shows the performance comparison against conventional analog MACs. Assuming that both MAC power and area linearly increases with the MAC resolution, the proposed PMAC achieves 48% higher and 26.6 times smaller normalized power efficiency and area, respectively compared to conventional arts. Compared to a DMAC synthesized in the same process, our PMAC achieves 8 times better power efficiency; by replacing the DMAC to a PMAC in well-structured DLAs, 66% power reduction will be achieved. The overheads are acceptable area increase and maximum operation speed limit of both only around 20%.

*Acknowledgements*

The authors would like to thank Xuan Yang, Edward Lee and Mark Horowitz for the valuable discussions on DLA architectures.